\documentclass[overfull,figures]{epl}

\usepackage[latin1]{inputenc}
\usepackage{amsmath,amssymb}
\usepackage{graphicx,textcomp}
\usepackage{dcolumn}
\usepackage{bm}

\title{Intercalation-enhanced electric polarization and chain formation of nano-layered particles}
\shorttitle{Electric polarization and chain formation of clay particles}

\author{J. O. Fossum\inst{1}  \and Y. M\'eheust \inst{1} \and K. P. S. Parmar\inst{1} \and K. D. Knudsen\inst{2} \and K. J. M{\aa}l{\o}y\inst{3} \and D. M. Fonseca\inst{1}}
\shortauthor{J. O. Fossum \etal}

\institute{
\inst{1} Department of Physics, NTNU, Hoegskoleringen 5, NO-7491 Trondheim, Norway\\
\inst{2}Physics Department, IFE, P.O. Box 40, NO-2027 Kjeller, Norway\\
\inst{3}Physics Department, UiO, Postboks 1048 Blindern NO-0316 Oslo, Norway
}

\graphicspath{
{./figures/},{./}
}


\begin{document}



\maketitle

\begin{abstract}
Microscopy observations show that suspensions of synthetic and natural nano-layered smectite clay particles submitted to a strong external electric field undergo a fast and extended structuring. 
This structuring results from the interaction between induced electric dipoles, and is only possible for particles with suitable polarization properties.
Smectite clay colloids are observed to be particularly suitable, in contrast to similar suspensions of a non-swelling clay. Synchrotron X-ray scattering experiments provide the orientation distributions for the particles. These distributions are understood in terms of competing (i) homogenizing entropy and (ii) interaction between the particles and the local electric field; they show that clay particles polarize along their silica sheet. Furthermore, a change in the platelet separation inside nano-layered particles occurs under application of the electric field, indicating that  intercalated ions and water molecules play a  role in their electric polarization. The resulting induced dipole is structurally attached to the particle, and this  causes particles to reorient and interact, resulting in the observed macroscopic structuring. The macroscopic properties of these electro-rheological smectite suspensions may be tuned by controlling the nature and quantity of the intercalated species, at the nanoscale.
\end{abstract}


In this letter we study colloidal suspensions of electrically-polarizable particles in non-conducting fluids. When such suspensions are subjected to an external electric field, usually of the order of 1kV/mm, the particles become polarized, and subsequent dipolar interactions are responsible for aggregating a series of interlinked particles that form chains and columns parallel to the applied field. This structuring occurs within seconds, and disappears almost instantly when the field is removed \cite{gastAdvCollIntSci,hillJAppPhys91,halseyScience92,halseySciAmer93,wenJAppPhys99}. It coincides with a drastic change in rheological properties (viscosity, yield stress, shear modulus, etc.) of the suspensions \cite{wenNatureMat2003}, which is why they are sometimes called electrorheological fluids (ERFs). This makes the mechanical behavior readily controllable by using an external electric field \cite{gastAdvCollIntSci,hillJAppPhys91,halseyScience92,halseySciAmer93,wenJAppPhys99,wenNatureMat2003,sprecherMatSciEng87}.  Particle size has a quite diverse impact on the behavior of ERFs \cite{tanPRE99}. The nature of the insulating fluid and of the colloidal particles determines the electrorheological behavior of the suspensions. The mechanism is not fully understood yet, but it is mainly triggered by the so-called interfacial polarization, and requires electric anisotropy of the particles \cite{haoLangmuir98}. Consequently, particle shape \cite{qiJPhysD2002} and surface properties \cite{durrschmidt_CollSurfA99} can also be critically important, as dielectric properties largely depend on them.

Clays as traditional material have played an important role throughout human history. Their common modern uses include nano-composites, rheology modification, catalysis, paper filling, oil well -drilling and -stability, etc \cite{veldeBOOK98}. Smectite (or 2:1) natural clay particles dispersed in salt solutions have been studied for decades \cite{veldeBOOK98}, and recently there has been a growing activity in the study of complex physical phenomena in synthetic smectites \cite{fossumNATO2000}. Much effort has gone into relating the lamellar microstructure of smectite clay-salt water suspensions to their collective interaction and to resulting macroscopic physical properties, such as phase behavior and rheological properties \cite{fossumNATO2000,mourchidLangmuir98,bonnLangmuir99,gabrielNature2001,diMasiPRE2001,lemaireEurophysLett2002,fossumENERGY2005}. Nematic liquid crystalline-like ordering in smectite systems have been characterized by the observation of birefringent domains with defect textures \cite{mourchidLangmuir98,gabrielNature2001,lemaireEurophysLett2002,fossumENERGY2005} or by Small-Angle X-Ray Scattering \cite{gabrielNature2001,lemaireEurophysLett2002}.

Wide Angle X-Ray Scattering (WAXS) studies of the well-characterized synthetic smectite clay fluorohectorite, in water suspensions \cite{diMasiPRE2001}, show that fluorohectorite particles suspended in water consist of "decks of cards" - like crystallites containing about one hundred 1 nanometer-thick platelets. Fluorohectorite has a rather large surface charge of 1.2 e-/unit cell, as compared to synthetic smectites such as laponite (0.4 e-/unit cell) \cite{kaviratnaJPhysChemSol96}; this high surface charge explains why clay stacks stay intact when suspended in water, unlike laponite or the natural clay montmorillonite. The natural smectite illite, on the other hand, behaves much like fluorohectorite in this respect \cite{brindleyBOOK1980}. One or more mono-layers of water may be intercalated into such "deck of cards" clay particles depending on temperature and relative humidity. For fluorohectorite, the dependence on those two parameters has already been mapped for hydration and dehydration by means of synchrotron X-ray scattering techniques \cite{daSilvaPRE2003}. However, the spatial configuration for the intercalated water molecules, with respect to the silica sheets and to the intercalated cations, is not precisely known yet. 
We show in this letter that the strong electro-rheological behavior exhibited by suspensions of smectite clay particles in silicon oil can be attributed to the intercalated species.

We studied four types of smectite suspensions in oil: Firstly three 
suspensions based on fluorohectorite (see \cite{diMasiPRE2001} for its origin and chemical formula), 
and secondly a mixed natural quick clay from the Trondheim region in Norway. 
The three types of fluorohectorite samples differ by the nature of the exchangeable cation, which is either mono- (Na$^+$), di- (Ni$^{2+}$) or tri-valent (Fe$^{3+}$).  
These synthetic samples are polydisperse with wide distributions of sizes (diameter up to a few micrometers, stack thicknesses around 100 nm) and aspect ratios. The quick clay is far less well-characterized than the synthetic clays, although from preliminary X-ray diffraction analysis we know that this mixed natural clay contains considerable amounts of illite and other natural smectite particles, in addition to the non-smectite clay kaolinite. The samples were initially prepared at ambient temperature by adding 1.5\% by weight of clay particles to the silicon oil Rotitherm M150 \cite{refOil} (viscosity 100cSt at 25{\textdegree}C). 
The experiments were performed using the supernatant of the oil suspensions, after sedimentation of the heavier particles.

\begin{figure}
\onefigure[width=0.5\textwidth]{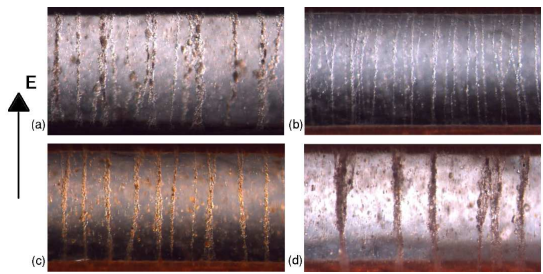}
\caption{\label{fig:microscope}  Microscope images of electrorheological chain formation in oil suspensions of smectite clays.
(a) Na-fluorohectorite  (b) Ni-fluorohectorite (c) Fe-fluorohectorite (d) Natural quick clay.}
\end{figure}

When placed between copper electrodes between which a sufficiently large electric field is applied (field strengths above 500 V/mm), the four types of samples exhibit the dipolar chain formation characteristic of electrorheological fluids (see Fig.~\ref{fig:microscope}). 
 The sample cell used for these observations consisted of two parallel and identical 1/2 mm thick copper electrodes separated by a gap of 2 mm and glued onto a transparent quartz glass microscope slide. The gap between the electrodes was closed at its ends by a non-conducting plastic material. The top part of the cell was open, and the sample cell was mounted horizontally, with the microscope slide flat down. A small volume ($<1$ ml) of the prepared sample was added and studied at ambient temperatures. The sample was illuminated from below, and observed from above in a stereomicroscope. An electric field  
 $E \sim 500$ V/mm was applied between the copper electrodes, and the changes in the sample were recorded by means of a digital camera connected to a PC. 
The process resulted in all clay particles being part of the electrorheological chain bundles after 10 to 20 s, and no motion being visible within the sample in less than 1 min. The critical electric field necessary to trigger the electrorheological behavior was found to be $E_c\simeq 400$ V/mm.

The procedure was then repeated using a suspension of kaolinite, a 1:1 natural clay which, in contrast to smectite clays, does not spontaneously intercalate cations and water molecules in-between its silica sheets. Kaolinite particles in their natural state have been reported in the literature to exhibit a weak electrorheological behavior when suspended in a silicon oil \cite{wangJMatChem2002}. Microscopy observations of kaolinite suspensions with the same density as the smectite suspensions exhibited electrorheology, but only  for $E\gtrsim 2 \un{kV/mm}$. 
They formed with characteristic time 10 to 100 times larger than that of the smectite suspensions. Furthermore, the bundle structure formed by the kaolinite suspensions appeared much less ordered than those observed with the smectites.

Relative orientations of the smectite particles inside the electrorheological chains were determined using synchrotron X-ray scattering experiments:
chain and column formations were observed by means of a video camera, while simultaneously recording X-rays scattered by the clay crystallites. These scattering experiments were performed at the Swiss-Norwegian Beamlines (SNBL) at ESRF (Grenoble, France), using the WAXS setup with a 2D mar345 detector at  beamline BM01A. 
\begin{figure}
\onefigure[width=0.50\textwidth]{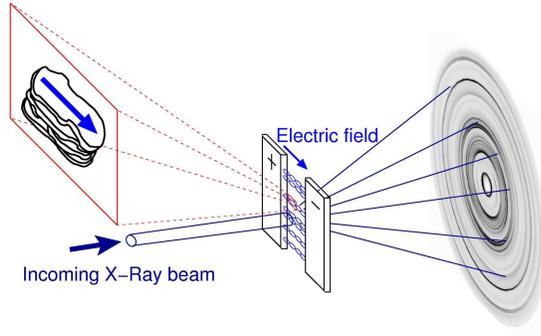}
\caption{\label{fig:experimental_sketch} Sketch of the synchrotron X-Ray scattering experiments. 
Corresponding diffractograms are shown in Figs.~\ref{fig:diffractograms}(b) and \ref{fig:diffractograms}(c). The magnified area shows a single nano-layered clay-particle inside a dipolar chain, with an arrow indicating the direction of the dipole moment induced by the external electric field.}
\end{figure}
 The sample cells used for these experiments differ from those described above in that the electrodes were placed vertically, the cells being closed at the bottom and with their top open,
which allows to partly fill the cell with sample from above (see a sketch of the experiment in Fig.~\ref{fig:experimental_sketch}). 

\begin{figure}
\onefigure[width=0.59\textwidth]{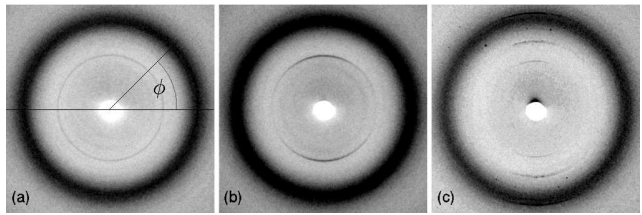}
\caption{\label{fig:diffractograms} Central regions of two-dimensional WAXS images of the electrorheological samples. (a) Suspension of Na-fluorohectorite at $E\sim 0$. (b) Suspension of Na-fluorohectorite for $E \sim 500$ V/mm $> E_c$. (c) Natural quick clay for $E \sim$ 500V/mm $> E_c$.}
 \end{figure} 
 Fig. \ref{fig:diffractograms} shows  three different two-dimensional diffractograms. Fig. \ref{fig:diffractograms}(a) is obtained from a suspension of Na-fluorohectorite prior to the application of an electric field. Each lamellar clay particle may be regarded as a single crystallite, and the particles are randomly oriented inside the sample, so the diffractogram is isotropic. The broad outermost ring is due to scattering from the silicon oil (characteristic length d $\sim$ 6.9-7.9 \AA), whereas the narrow symmetric ring at lower scattering angles is the (001) Bragg peak from the lamellar clay stacks with 1 water layer intercalated (d $\sim$ 12.3 \AA). In the presence of an electric field (Fig.~\ref{fig:diffractograms}(b)), in contrast, the (001) Bragg peak has become anisotropic due to particle orientation in the field.
 In addition, the number of water layers intercalated was determined directly from the positions of the Bragg peaks in reciprocal space (as in \cite{daSilvaPRE2002}). Fig. \ref{fig:diffractograms}(c) shows the diffractogram of a suspension of natural quick clay, for $E>E_c$. 
 Three anisotropic scattering rings are visible: the 1st order of illite (d $\sim$ 10.1 \AA), and the 1st (d $\sim$ 14.28 \AA) and 2nd order of an another smectite clay. The whole diffractogram (not shown here) also reveals the 3rd order ring of illite, and the 3rd and 4th order rings of the other smectite clay. All the visible diffraction rings correspond to diffraction by smectite clays and are anisotropic.
\begin{figure}
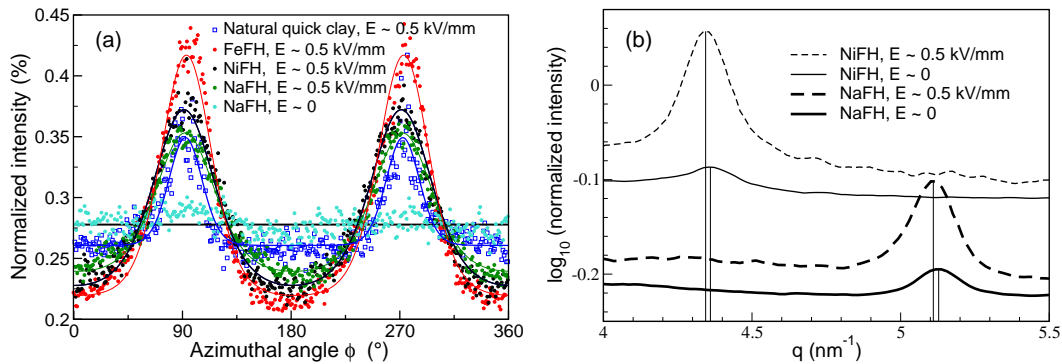

\twoimages[width=0.48\textwidth]{figure4.eps}{figure5.eps}
\caption{\label{fig:azim_plots}(a) Dependence of the intensity of circular scattering rings on the azimuthal angle $\phi$. (b) Scattering lines along the $q$ direction for the Na- and Ni- fluorohectorite samples in the 1WL hydration state (corresponding respectively to characteristic spacings $d=2 \pi / q=12.26$  {\AA} and $d=14.41$ {\AA} for $E~\sim 0$): a systematic increase of the platelet separation inside the clay particles occurs under application of the electric field, and results in a shift of the corresponding diffraction peak. The spectra have been translated vertically with respect to each other, for clarity.}
 \end{figure}
 Fig.~\ref{fig:azim_plots}(a) shows how the intensity of circular scattering rings such as those presented in Fig.~\ref{fig:diffractograms} evolve as a function of the azimuthal angle, between 0 and 360{\textdegree}.
In the case of fluorohectorite, we have considered the ring at a radial position corresponding to the first order Bragg peak for the clay stack \cite{daSilvaPRE2002}, which is characteristic of 1 mono-layer of intercalated water at ambient temperatures ($d \sim 12.4$ {\AA} for the Na-fluorohectorite).
For the natural sample, we have considered the lowest order Bragg peak of the smectite, illite (see Fig.~\ref{fig:diffractograms}(c)). The scattered intensity at a given azimuthal angle is proportional to the number of particles that meet the Bragg condition for that angle, which implies that the shapes of the scattered intensities in Fig.~\ref{fig:azim_plots}(a) provide the orientation distributions of clay particle orientations inside the chains and columnar structures. 
For $E\sim 0$, the intensities are independent of the azimuthal angle (except for the experimental noise). 
For $E>E_c$, the azimuthal positions of the maxima along the plots in Fig.~\ref{fig:azim_plots}(a) demonstrate that the preferred orientation of the clay particles is with the lamellar stacking plane parallel to the direction of the electric field.
Since the collective dipolar interactions between particles align their dipole moments with the electric field, the preferred direction of polarization for each stacked particle could also be inferred, as indicated in the insert of Fig.~\ref{fig:experimental_sketch}: it is parallel to  the silica sheets.

Colloids that make "good" electrorheological particles are known to possess two important properties, which allow them to not only polarize, but also to rotate, under the effect of the electric field: (i) their interfacial polarizability is important, and (ii) they are electrically anisotropic  \cite{haoLangmuir98,haoAdvColIntSci2002}. In the case of our platelet-shaped nano-layered smectite clay particles, particle polarization occurs along their silica sheet. A fine monitoring of the scattering line along $q$ shows that the application of the electric field results in a subtle increase in the characteristic separation between adjacent platelets inside clay particles (see Fig.~\ref{fig:azim_plots}(b)), of about $0.05$ {\AA}. This is consistent with polarization occurring mainly by movement of charges along the quasi-2D inter-layer space, perpendicular to the stacking direction, by movement of ions out of preferred sites on inter-layer surfaces. We believe that the intercalated cations and possibly water molecules dominate the interfacial polarization process for smectite particles, which explains the low critical field and fast response observed for the smectite suspensions as compared to the kaolinite suspensions.

 Considering that the induced dipole is structurally forced to remain in the plane of the silica sheets, we can explain why the particles rotate under application of the field, and we can further describe the distribution of particle orientations around the mean orientation
in terms of a competition between (i) the homogenizing entropy and (ii) the aligning effect of the strong external electric field.
Indeed, the Gibbs energy of the colloid population can be written as the sum of an entropic term and of the interaction energy of the clay particles with a local mean electric field aligned with the external field.
These two terms are dependent on the functional form of the orientation distribution of the particles, $f$. Minimizing the Gibbs energy with 
respect to that functional form yields the orientation distribution at equilibrium. This calculation is presented in details in a separate manuscript \cite{MeheustUnpub}; it follows the line of the well-known
theory developed by Maier and Saupe to describe a different geometry, namely nematic ordering \cite{MaierNaturforsch58,MaierNaturforsch59}). In our system, the interaction energy is different from that of Maier and Saupe, but the final functional form obtained for $f$ is identical:  $f(\theta)\propto \exp (m\, \cos \theta)$, where the angle $\theta$ denotes the deviation from the mean orientation, while the physical parameters characteristic of the system are contained in the expression of the parameter $m$ \cite{MeheustUnpub}. Obviously, the expression for $m$ is different in our system from what it is in the Maier-Saupe
theory. For each type of clay, we have fitted a Maier-Saupe profile to the data in order to quantify the orientational ordering of particles within the chains and the columns from an estimate of the distribution width (see fully drawn lines onto the scatter plots for the data in Fig.~\ref{fig:azim_plots}). The half widths at half maximum are: $25.1 \pm 0.3${\textdegree} for Na-fluorohectorite, $25.6 \pm 0.3${\textdegree} for Ni-fluorohectorite, $22.4\pm0.3${\textdegree} for Fe-fluorohectorite, and $19.9\pm0.7${\textdegree} for the natural quick clay, which corresponds to values of the corresponding order parameter, 
$S=\langle 3 \sin^2 \theta -2 \rangle_f$\cite{MeheustUnpub}, of  about 0.74-0.79 (fluorohectorite samples) and 0.83 respectively \cite{MeheustUnpub}.
This difference in order parameter magnitude may be related to differences in particle size and polydispersity.

\begin{figure}
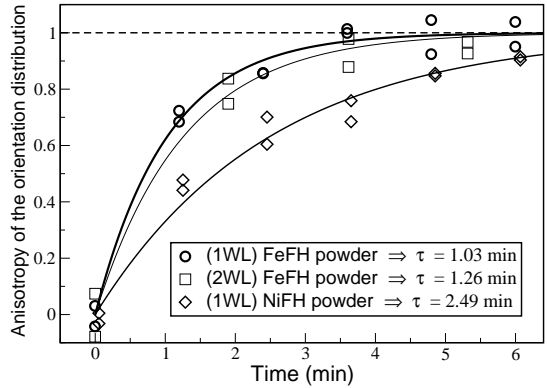

\onefigure[width=0.5\textwidth]{figure6.eps}
\caption{\label{fig:dynamics_wet_and_dry} Anisotropy of the orientation distribution as a function of time during the buildup of the electrorheological structure. The anisotropy is computed as the amplitude of the 
oscillations in Fig.~\ref{fig:azim_plots}(a), normalized so that this amplitude be equal to $1$ at large times. A characteristic time $\tau$ for particle orientations is obtained by fitting an exponential curve to the experimental data.}
 \end{figure}
What exactly occurs in the between-platelets space under the application of the electric field remains unclear.
We have monitored the transient buildup of the orientation distributions (as represented in Fig.~\ref{fig:azim_plots}(a)) 
following the sudden switch-on of the electric field above $E_c$. Fig.~\ref{fig:dynamics_wet_and_dry} shows the amplitude of the azimuthal profiles (such as those shown in Fig.~\ref{fig:azim_plots}(a)) as a function of time; the plots have been rescaled so that the amplitude at infinite time be equal to $1$. Characteristic time scales $\tau$ were obtained from fits in the form $1-\exp(-t/\tau)$; the time scales obtained are of the same order as or a bit larger than the durations after which no particle movement is visible in the microscope. 
As shown by Fig.\ref{fig:dynamics_wet_and_dry}, those dynamical experiments did resolve clear differences in characteristic time scales for chain ordering depending on the type of intercalated ion; for example, a factor of 2 (faster dynamics) was observed in the buildup velocity for Ni-, as compared to Fe-, fluorohectorite samples. In contrast, when preparing fluorohectorite samples in different hydration states, we were not able to clearly resolve a dependency of the transient dynamics on the number of intercalated water layers, although we cannot exclude the possibility of such effects existing within our experimental uncertainty. Yet, our present data clearly indicates that the role of ions in the polarization process dominates that of water molecules. 
Understanding the mechanisms underlying the particle polarization shall require experiments utilizing other experimental techniques. An EXAFS study of the vicinity of the intercalated Nickel cations during particle polarization, which can show how their spatial configuration inside the interlayer space is modified by the electric field, is planned. Note that for these suspensions, one may be able to control the characteristic time for dipolar chain formation on the macro-scale by manipulating, on the nano-scale, the nature of the intercalated ions as well as the number of intercalated water layers. Among foreseeable applications are the ones common to electrorheological fluids \cite{gandhiBOOK}. Another potential application is a method for separating smectite from non-smectite components, in natural clays.

\acknowledgements
We gratefully acknowledge assistance from the staff of the Swiss-Norwegian Beam Lines at ESRF. We also thank Jan J{\o}nland for providing us with the natural clay. This work was supported by the Research Council of Norway (RCN) through the NANOMAT Program: RCN project numbers 152426/431, 154059/420 and 148865/432, as well as 138368/V30 and SUP154059/420.

\end{document}